\documentstyle[12pt,amsmath]{article}
\newcommand{\De}{\Delta}
\newcommand{\ph}{\varphi}
\renewcommand{\t}{\theta}
\begin{document}
\begin{center}
{\Large\bf On some inverse problems in nuclear physics}
\vskip 5mm
B.Z. Belashev${^{1)}}$ and M.K. Suleymanov${^{2,3)}}$
\vskip 5mm

{\small ${^{1)}}$ Institute of Geology, Karelian Research Centre, RAS, Petrozavodsk, Russia\\
${^{2)}}$ LHE JINR, Dubna, Moscow region, Russia\\
${^{3)}}$ NPI ASCR, Rez near Prague, Czech Republic}
\end{center}
\vskip 5mm
\centerline{\bf Abstract}

Some inverse problems in high-energy physics, neutron dif\/fraction and NMR 
spectroscopy are discussed. To solve them, the Fourier integrated 
transformation method and the Maximum Entropy Technique (MENT) were used. 
The integrated images of experimental distributions are shown to be 
informative when determining the space-time parameters of a particle 
generation zone and when analysing blurred spectra. The ef\/f\/iciency of the 
above methods was checked by comparing relevant results with the results 
obtained independently.
\medskip

An inverse problem arises where the information required can not be obtained 
through direct measurements. The majority of inverse problems in nuclear 
physics is connected to interpretation of experimental data on scattering 
high-energy particles on targets. The scattering pattern is used to determine 
the structure of colliding particles, interaction potential and the dynamics 
of processes occurring upon collision. In some cases, a nucleus or a particle 
is understood as a probe with known magnetic moment, wave length and energy 
level system. Its interaction with a substance and a f\/ield is used to assess 
the concentration of the substance, the characteristics of the medium or the 
conf\/iguration of the f\/ield. Such inverse tasks are typical of NMR and NGR 
spectroscopy, neutron dif\/fraction, activating analysis, emitted tomography 
and other applications of nuclear physics. 

In this paper, some inverse problems arising in experiments in high-energy 
physics are discussed~\cite{[1],[2],[3],[4],[5]}. To solve them, f\/iltration methods are used. 
Experimental data are subjected to Fourier transformation to obtain a result 
without a substantial use of a priori restrictions~\cite{[3],[6]}. This property is 
especially important when selecting models of processes in cases when the 
experimental distributions are compatible with several models and the 
application of approximation methods is inef\/fective.

One question, which arises when solving an inverse task, is whether the 
results obtained are proven. This question can be answered by comparing the 
results obtained by independent methods. To illustrate such a comparison, 
dif\/fusion spectra were used as an example, their resolution being increased 
instrumentally and a posteriori through the Maximum Entropy Technique (ÌENT)%
~\cite{[7],[8],[9]}.

The goal of our study was to show that the same type of mathematical and 
algorithmic approaches, based on integrated transformations, can be used to 
solve some inverse tasks in nuclear physics.
\bigskip

\centerline{SPACE-TIME CHARACTERISTICS OF THE  MULTIPLE  BIRTH OF PARTICLES}
\medskip

Using Podgoretsky and Kopylov's approach, light can be thrown on the 
space-time characteristics of the multiple birth of particles by studying 
correlations of identical particles with close 4-impulses~\cite{[1]}. 
The probability $W(q_0,\Vec{q})$ of detecting two identical particles 
with 4-impulses $P_1=\{\varepsilon_1,\Vec{p_1}\}$  and
$P_2=\{\varepsilon_2,\Vec{p_2}\}$, emitted by two point sources with 
the coordinates $\{t_1,\Vec{r_1}\}$ and $\{t_2,\Vec{r_2}\}$, is expressed 
by the formula:
\begin{equation}
W(q_0,\Vec{q})=1+\frac{\cos[\Vec{q}(\Vec{r_1 }-\Vec{r_2 })-q_0(t_1-t_2 )]}%
{1+(q_0\tau)^2}
\end{equation}
where $\Vec{q}=\Vec{p_1}-\Vec{p_2}$, $q_0=\varepsilon_1-\varepsilon_2$, $\tau$
is the lifetime of the sources. For identical particles of identical energy
$q_0=0$, a relationship (1) is determined by a spatial variable
$\Vec{R}=\Vec{r_1}-\Vec{r_2}$:	
\begin{equation}
W(\Vec{q})=1+\cos(\Vec{q}\Vec{R}).
\end{equation}
The above formulas hold for identical particles that obey Boze-Einstein 
statistics, e.g. the $\pi$-mesons of large impulses.
Let us assume that $f(\Vec{R})$ is the normalized spatial distribution of 
the sources of particles on the $\Vec{R}$. Then the formula (2) for the 
probability $W(\Vec{q})$ has the form:
\begin{equation}
W(\Vec{q})=1+\int f(\Vec{R})\cos(\Vec{q}\Vec{R})d\Vec{R}.
\end{equation}
In this case $f(\Vec{R})=f(-\Vec{R})$, because two identical particles are 
indexed by two equivalent methods. The function $W(\Vec{q})$
does not depend on the sign $\Vec{R}$ and the integrated addend in (3) is 
the Fourier image $F(\Vec{q})$ of the function $f(\Vec{R})$
\begin{equation}
W(\Vec{q})=1+F(\Vec{q}).
\end{equation}
Thus, the experimental relationship $W(\Vec{q})$ is used to obtain the 
Fourier image of the distribution function $f(\Vec{R})$
\begin{equation}
F(\Vec{q})=W(\Vec{q})-1
\end{equation}
and to use it, in turn, to calculate the distribution function $f(\Vec{R})$
by inverse Fourier transformation.

In collisions of identical particles, experimental data can directly provide 
information on the distribution density of the sources $\rho(\Vec{r})$
in the generation f\/ield, provided the sources of the particles are independent%
~\cite{[2]}. The Fourier image $G(\Vec{q})$ of the function $\rho(\Vec{r})$
is calculated using the formula                  
\begin{equation}
G(\Vec{q})=\sqrt{W(\Vec{q})-1}
\end{equation}
and the function $\rho(\Vec{r})$ is estimated by its inverse Fourier transformation.
The three-dimensional function $\rho(\Vec{r})$ can be represented for 
convenience as cutting of the generation zone by a system of parallel planes 
or a projection of the generation zone to a plane perpendicular to the preset 
direction. The condition $\rho(\Vec{r})\ge 0$ restricts experimental 
distributions to a certain class of functions. Such a restriction is 
automatically removed if the distribution of the sources is reconstructed 
by its Fourier image using ÌENT~\cite{[10]}.

A similar technique of selecting identical particles with $\Vec{q}=0$
to estimate the distribution function of the sources on the variable
$t=t_1-t_2$ is impossible as it results in the value $q_0=0$, which 
follows from a condition relating the masses, energies and impulses of 
identical particles. When discussing a general case, we will not restrict the
$q_0$ value. Let us introduce the normalized distribution function
$f_1(\Vec{R},t,\tau)$ and re-write the formula (1), considering a 
trigonometrical identity:
\begin{equation}
\cos(\Vec{q}\Vec{R}-q_0t)=\cos(\Vec{q}\Vec{R})\cos(q_0t)+\sin(\Vec{q}\Vec{R})\sin(q_0t)
\end{equation}
as 
\begin{multline}
W(q_0,\Vec{q})=1+\int f_1(\Vec{R},t,\tau)\frac{\cos(\Vec{q}\Vec{R})%
\cos(q_0t)d\Vec{R}dtd\tau}{1+(q_0\tau)^2}+\\
+\int f_1(\Vec{R},t,\tau)\frac{\sin(\Vec{q}\Vec{R})\sin(q_0t)d\Vec{R}dtd\tau}%
{1+(q_0\tau)^2}.
\end{multline}
The second integral addend in the formula (8) is equal to zero because the 
methods of indexing of a particle in a pair are indentical, the function
$f_1(\Vec{R},t,\tau)$ is even and the function of a sine on the variables
$\Vec{R}$ and $t$ is odd.

In the assumption about the independence of spatial, temporal and relaxation 
distributions, described accordingly by the normalized functions
$f(\Vec{R})$, $\chi(t)$ and $\ph(\tau)$, expressed by the condition
\begin{equation}
f_1(\Vec{R},t,\tau)=f(\Vec{R})\cdot\chi(t)\cdot\varphi(\tau)
\end{equation}
formula (8) is re-written as
\begin{equation}
W(q_0,\Vec{q})=1+\int f(\Vec{R})\cos(\Vec{q}\Vec{R})d\Vec{R}%
\int\chi(t)\cos(q_0t)dt\int\frac{\varphi(\tau)d\tau}{1+(q_0\tau )^2}.
\end{equation}
The f\/irst two integrated co-multipliers in (10) represent the Fourier images
$F(\Vec{q})$ and $X(q_0)$ of the normalized spatial $f(\Vec{R})$
and temporal $\chi(t)$ distributions. The third integrated multiplier does 
not depend on the magnitude of $q_0$, which follows from the condition 
of normalizing of the function $\ph(\tau)$
\begin{equation}
\int\limits_0^\infty\varphi(\tau)d\tau=\int\limits_0^\infty\varphi(q_0\tau)d(q_0\tau)=1.
\end{equation}
and can be written as constant
$\displaystyle C=\int\limits_0^\infty\frac{\varphi(q_0\tau)d(q_0\tau)}{1+(q_0\tau)^2}>0$.
The formula (10) then has the form                                     
\begin{equation}
W(q_0,\Vec{q})=1+CF(\Vec{q})X(q_0)
\end{equation}
can be used to estimate the function $CX(q_0)$ from the experimental relations
$W(q_0,\Vec{q})$ and $W(0,\Vec{q})$
\begin{equation}
CX(q_0)=\frac{W(q_0,\Vec{q})-1}{F(\Vec{q})}=\frac{W(q_0,\Vec{q})-1}{W(0,\Vec{q})-1}.
\end{equation}
The distribution function of the sources $\chi(t)$ is determined by inverse 
Fourier transformation of the function $CX(q_0)$ and subsequent normalization 
by unit.

In this problem, the integrated images of spatial and temporal distribution 
of the sources of particles are present in the data obtained by the 
interference experiment, and can be revealed if the data are represented in 
a special manner. In some cases, an artif\/icial transition from experimental 
distributions to their integrated images and nonlinear operations with them 
can be useful. There is a class of such operations that can be done in 
the case of inverse transition to obtain more informative estimates of 
data than in initial distributions~\cite{[11]}.
\bigskip

\centerline{RETRIEVAL OF BARION RESONANCES IN EFFECTIVE MASS SPECTRA}
\medskip

The aim of our study was to investigate two-particle correlations of 
the reaction products of  multiple birth at high energies that are 
observed in the distributions of pairs of secondary particles on the 
ef\/fective mass $M_{eff}$ and provide information on compound systems 
or resonances formed during particle generation. Of interest for the 
study of collision dynamics are $\Delta$-isobars, barion resonances 
that form a connected short-lived system from a nucleor and a $\pi$-meson. 
This interest is due to the fact that the birth of such resonances 
on nuclei upon collision of hadrons with the nuclei at low energies 
can provide the main channel for elastic scattering. Isobars can give 
information about interaction with other fragments of the nucleus and 
enable us to experimentally study the colour degrees of freedom.

The purpose of the study of the ef\/fective mass spectra $(\pi^\pm p)$
pairs in $\pi^-p$ and $\pi^{-12}C$ interactions at $P_{\pi^-}=40GeV/c$
is to determine the structure of the spectra and to relate the birth 
of barion resonances and a special case of the multiple generation of 
hadrons, such as birth of cumulative particles in $\pi^{-12}C$
interactions~\cite{[4]}.

The spectra of the ef\/fective masses of pairs were plotted on the basis of 
non-elastic 11688 $\pi^-p$ and 8642 $\pi^{-12}C$ interactions obtained by 
analysis of stereo photos taken by a two-metre propane camera of LHE JINR 
irradiated by a bundle of of $\pi$-mesons from the Serpukhov accelerator~\cite{[12]}.
The ef\/fective mass of pairs was calculated using the formula:
\begin{equation}
M_{eff}=\sqrt{m_p^2+m_\pi^2+2(E_pE_\pi-p_pp_\pi\cos\t)}
\end{equation}
using the masses of proton and $\pi$-meson $m_p$ and $m_\pi$ their total 
energies $E_p$ and $E_\pi$ the absolute values of impulses $p_p$ and $p_\pi$
and an angle $\t$ between impulses (in a laboratory system). The impulse 
of the protons selected varied from 140 to 700 MeV/s.

To establish a relation between the birth of $\De$ isobars and the cumulative 
irradiation of $\pi$-mesons in interactions, two schemes of selection of 
secondary  $\pi$-mesons in the event were used. In the f\/irst scheme, all 
$\pi$-mesons of the event were considered. In the second scheme, $\pi$-meson 
with a maximum cumulative order value in the event was excluded. The 
cumulative order of $\pi$-meson was determined in accordance with the formula
$\displaystyle n_c=\frac{E-p_{//}}{m_N}$ through total energy $E$, longitudinal 
impulse $p_{//}$ of $\pi$-meson and the mass of the nucleor $m_N$.
The structure of the ef\/fective mass spectra $\pi^\pm p$
pairs was analysed using two data processing methods based on Fourier 
transformation. In the former method, we revealed the periodicities of 
the dependence of the module of the Fourier image of a symmetric spectrum on 
angular frequency. These, in turn, were used to determine the positions of 
structural components corresponding to resonance masses~\cite{[3]}. The results 
obtained by this method are shown in Figure 1. The positions of the 
specif\/ied characteristics of the spectra are shown by arrows, position 
errors by segments above the arrows and the value of the mass of known 
resonances by a number.

In the latter method, the idea to control the widths of the components of 
the ef\/fective mass spectrum was used~\cite{[6]}. The widths of the components in 
the estimation of the spectrum were diminished by multiplying the Fourier 
image of the spectrum by an exponential function. Upon inverse Fourier 
transformation of such an image, we the resultant estimate had a higher 
resolution than the initial spectrum and exhibited some characteristics that 
could be compared in ef\/fective mass to barion resonances~\cite{[4]}.

The results obtained by this method are presented in Figure 2 which shows 
experimental distributions $\displaystyle\frac{dN(M_{eff})}{dM_{eff}}$
and calculated estimates of $f(M_{eff})$ coresponding to the pairs
$\pi^\pm p$ in the former scheme of selection of $\pi$-mesons of the 
event (a and b), and $\pi^\pm p$ pairs in the latter scheme of selection 
(c and d). The main peak of the structures detected corresponds to a mass 
of 1.232 MeV/c2 and additional peak corresponds to mass 1.650 MeV/c2. 
When cumulative $\pi$-mesons were eliminated, the estimates of the ef\/fective 
mass spectrum declined in a uniform manner by ca. 30\% without changing the form.

The characteristics obtained in the ef\/fective mass spectra of $\pi^\pm p$
pairs were compared to $\Delta^{+_0+}(1.232)$, $\Delta^{+_0+}(1.650)$,
$\Delta^{+_0+}(1.670)$, $\Delta^{+_0+}(1.910)$ isobars in $\pi^-p$
interactions and to $\Delta^{+_0+}(1.232)$, $\Delta^{+_0+}(1.650)$
isobars in $\pi^{-12}C$ interactions. Analysis of the average multiplicity 
of $\pi^+$ and $\pi^-$-mesons in $\pi^{-12}C$ interactions at 
$P_{\pi^-}=40$ GeV/c, made with regard for the background of the spectra 
has shown that elimination of any $\pi^-$ meson in the event decreases by 
about 30\% in the distributions $\frac{dN(M_{eff})}{dM_{eff}}$. 
This result is believed to show that the generation of cumulative $\pi$-mesons 
and the birth of barion resonances are two independent processes.
\bigskip

\centerline{RAPIDITY-BASED PARTICLE DISTRIBUTION STRUCTURE}
\medskip

According to modern concepts of interaction between high-energy hadrons 
and nucleors and nuclei, secondary particles dominantly arise during the 
hadron transformation of fast quark parton into a jet of particles~\cite{[13]}. It 
is convenient to study this jet hadron transformation, seen in short-range 
correlations of interaction products, in the rapidity-based distributions 
of reaction products, observing the grouping of particles relative to 
several characteristic rapidities.

Rapidity is determine as $y=(1/2)\ln((E+p_{//}/(E-p_{//}))$ where
$Å$ and $p_{//}$ are energy and the longitudinal impulse of a secondary particle.

Rapidity-based distributions are typically plotted as smooth curves 
without visible characteristics, which makes it dif\/f\/icult to select jets. 
To detect the characteristic rapidities of the multiple generation of
$\pi^\pm$-mesons in $\pi^-p$ and $\pi^{-12}C$ reactions at $P_{\pi^-}=40$
GeV/c the method, used earlier to decrease the widths of distribution 
components, was applied~\cite{[7]}.

Input data and the results of the application of this method are shown 
in Figure 3. The rapidity-based distributions of $\pi^\pm$-mesons in these 
reactions (a and b) are highly blurred, but their contrasting estimations 
(c and d) are structurally complex. In $\pi^-p$ interaction, the estimate 
of rapidity-based distribution of $\pi^+$-mesons is a doublet with peaks 
$y\approx 1.4$ and $y\approx 2.8$. For $\pi^-$-mesons, it consist of three 
peaks with $y\approx 1.2$, $y\approx 2.8$, $y\approx 4.7$. The estimates 
of rapidity-based distribution in $\pi^{-12}C$ interactions for $\pi^+$-mesons 
contains three peaks with rapidities $y\approx 1.4$, $y\approx 2.8$, 
$y\approx 3.3$ and three peaks with rapidities $y\approx 0.9$, $y\approx 2.5$,
$y\approx 4.4$ for $\pi^-$-mesons with a highly dif\/fused second peak in the 
rapidity interval of 1.8--3.2.

The results obtained can be used to compare the characteristic rapidities 
of secondary $\pi$-mesons over the ranges 1.0--1.6; 1.8--3.2; 4.0--4.4 
to the areas of fragmentation of a target, a colliding $\pi$-meson. The 
number and values of the characteristic rapidities of the process depend 
on the mass of the target and the charges of secondary $\pi$-mesons.
\bigskip

\centerline{INCREASING THE INFORMATIVE CAPACITY OF SPECTRA BY MENT}
\medskip
A solution can be increased by dif\/fusive spectra by the MENT. To increase 
the solution of experimental spectra, a variant of ÌENT with trial 
dif\/fusion functions is used~\cite{[7]}. This variant is similar to Fourier 
method for regulating distribution widths, but has the best solution and 
stability to the noise~\cite{[11]}.

In this study, MENT was used to reconstruct the inner structure of dif\/fusive 
neutron diagrams and the NMR spectra of solid state. These spectra typically 
provide a tool for recording spectra that dif\/fer in resolution. This 
allows to compare the results obtained by processing the original dif\/fuse 
low-resolution spectra with those obtained by recording high-resolution 
spectra and to f\/ind out whether dif\/fuse spectra can be processed.

Neutron dif\/fraction is an ef\/fective tool for the structural study of a 
substance due to various properties of a neutron and the comparability of 
its wave lengths and energies of neutrons to interatomic distances and 
excitation energies in solid and liquid states.

A convenient scheme of neutron dif\/fraction, compatible with impulse reactors 
of neutrons, is the reflection of neutrons from an atomic plane at a 
constant angle with decomposition of reflection intensity in a spectrum 
on the wave length of neutrons or the flight time by neutrons of the 
def\/ined base. When developing this technique, Fourier dif\/fraction was 
accomplished with the help of a scheme of Fourier breakers of a beam on 
impulse reactors of neutrons. Neutron Fourier dif\/fraction is characterized 
by a high spectral solution and by smaller level of a hum.

Fig. 4 shows neutron diagram of powder $YBa_2(Cu,Fe)_3O_{63}Y_{12}$
registered with usual solution by a time of flight dif\/fractometer (a) and 
neutron diagram of this sample measured on the IFR-2 reactor (LNF JINR 
Dubna) with high solution by Fourier dif\/fractometer (b)~\cite{[8]}. Outcome of 
treatment of diagram (a) by the MENT: estimation and its touch diagram 
are gave on (c).

The comparison of these graphs suggests the similarity of structures 
represented by the MENT touch diagram and Fourier diagram.

NMR spectra provide important chemical, biophysical and medical information.

Magnetic resonance is characteristic of nuclei with a spin dif\/fering from 
zero placed in a magnetic f\/ield. It is observed as the precession of 
the magnetic moments of the nucleus in a magnetic f\/ield. The macroscopic 
vector of magnetization can be recorded during the interaction of moments 
with electromagnetic radiation containing frequency components near Larmor 
frequency $\omega_0$ or under the influence of fast changes in the 
orientation of an external magnetic f\/ield. For this purpose, in 
addition to a static magnetic f\/ield, a radio frequency f\/ield with Larmor 
frequency $\omega_0$ is applied perpendicularly to direction of static 
f\/ield or several types of RF-impulses (called $\pi/2$, $\pi$, $3\pi/2$ and
$2\pi$ in accordance with the angle of turn of the magnetization 
vector in a rotating coordinate system) are used to initiate transition 
processes in a system of nuclear spins. A signal of free collapse 
of induction after the cessation of the action of $\pi/2$ of impulse 
can be used, as a result of inverse transformation of Fourier, 
to obtain data on the characteristic frequencies of a system of 
nuclear spins as well as spin-lattice and spin-spin relaxation times.

NMR spectroscopy can conveniently be applied to liquids the spectra of 
which are represented by narrow spectral lines. The NMR spectra of 
solids, recorded by the forced precession method, are, on the contrary, 
dif\/fuse and are hard to interpret. The reason for that is that the 
averaging of dipole-dipole spin-spin interaction is less ef\/f\/icient 
than in liquids. Artif\/icial averaging became possible after using 
special sequences of $\pi/2$ and $\pi$ impulses~\cite{[9]}. The NMR spectra 
of solids then became similar to those of liquids.

Applying MENT to a low-resolution NMR, we can compare the estimate 
of the spectrum obtained with a high-resolution NMR spectrum. 
In this example, our purpose was to check up the ability of the 
MENT algorithm for the dif\/fuse spectra of solids that have a constant 
noise level. 

Figure 5 shows the NMR spectrum ${}^{13}C$ of adamantane, an organic compound (a)
recorded by the forced precession method (curve 1) and with the help $\pi/2$,
$\pi$ sequences of impulses (curve 2)~\cite{[9]}. Also shown in the f\/igure is a 
MENT estimate (curve 3) with its touch diagram (curve 4) (b) obtained by 
processing the dif\/fuse NMR spectrum.

Characteristic of a processed spectrum is its arbitrary position on the 
ordinate axis, i.\,e. the lack of information on the constant component 
of hum noise. It is usually dif\/f\/icult to process and interpret such 
spectrum. Indeed, the touch diagram of the MENT estimate of this 
spectrum shows a series of peaks mot of which have a background 
origin. However, there are typically two intensive peaks that reflect 
the true structure of the high-resolution NMR spectrum recorded.

This example shows that the hum noise of a dif\/fusion spectrum should be 
accurately estimated to draw correct conclusions about its structure. 
It also shows that the position of the most intensive components of a 
dif\/fusion spectrum can be determined even though there is no information on 
the constant constituent of hum noise.

\bigskip\noindent Conclusions:\medskip

1. Based on the interference approach, a method for reconstruction of the 
space-time characteristics of a particle generation zone in collisions 
of high-energy particles is proposed. This method can be employed to determine 
the density functions of the space-time distributions of the sources of 
particles from the experimental distribution of pairs of secondary identical 
particles with respect to a dif\/ference in their impulses and energies, thus 
solving an inverse problem. The statistical provision of modern experiments 
of high-energy physics favours the application of this method.

2. The application of the methods of Fourier analysis of dif\/fusion spectra 
to search for barion resonances and characteristic rapidities in the reactions
of $\pi^-p$ and $\pi^{-12}C$ interactions at $P_{\pi^-}=40$ GeV/c has made it 
possible to elucidate the structural characteristics of experimental distributions and 
to compare them with known $\De$ isobars and zones of fragmentation of reaction 
products in the rapidity space. The results obtained by analysing the spectra 
of the ef\/fective masses of pairs using various schemes of selection of 
$\pi$-mesons in an event show that the formation of $\De$ isobars and the 
emission of cumulative $\pi$-mesons are independent processes. This 
approach can be used to search for various resonances, multi-quark  systems, 
exotic mesons and purely glue states~\cite{[13]}.

3. In inverse problems arising in applied nuclear physics in 
neutron dif\/fraction and NMR, MENT-based methods of analysis of dif\/fuse 
spectra  gave results comparable to those obtained by high-resolution 
instrumental methods. A high informative capacity of the MENT methods 
was supported by processing simple and complex spectra in the absence 
of data on the constant constituent of hum noise.

The authors wish to thank S.~S.~Shimansky for useful discussions.

\listoffigures
\begin{itemize}
\item[Fig.1.] Results of the application of the modif\/ied Fourier 
algorithm to ef\/fective mass spectra of ($\pi^\pm p$) pairs in $\pi^-p$
and $\pi^{-12}C$ interactions at $P_{\pi^-}=40$ GeV/c.
\item[Fig.2.] Ef\/fective mass spectra $dN(M_{eff})/dM_{eff}$ 
of ($\pi^\pm p$) pairs ($\circ$), their estimation $f(M_{eff})$ ($\bullet$) 
for two schemes of selection of $\pi$-mesons.
\item[Fig.3.] Rapidity-based distribution of $\pi$-mesons in $\pi^-p$ (à)
and $\pi^{-12}C$ (b) interactions and their estimates (c) and (d), respectively, 
obtained with the help of Fourier algorithm used to control spectral line widths.
\item[Fig.4.] Neutron diagram of a powder combination  measured with 
a usual time flight dif\/frac\-tometer (a) and by Fourier dif\/fractometer under 
high-solution conditions on reactor IFR-2 (JINR, DUBNA) (b) and obtained as 
a result of MENT processing of the diagram (a): the MENT estimation and its 
touch-diagram (c).
\item[Fig.5.] NMR spectrum ${}^{13}C$ of adamantane recorded by the 
forced precession method (curve 1) and with the help $\pi/2$, $\pi$-sequence 
of impulses (curve 2) (a) and the MENT estimation of a curve 1 (curve 3) 
and its touch diagram (curve 4) (b).
\end{itemize}
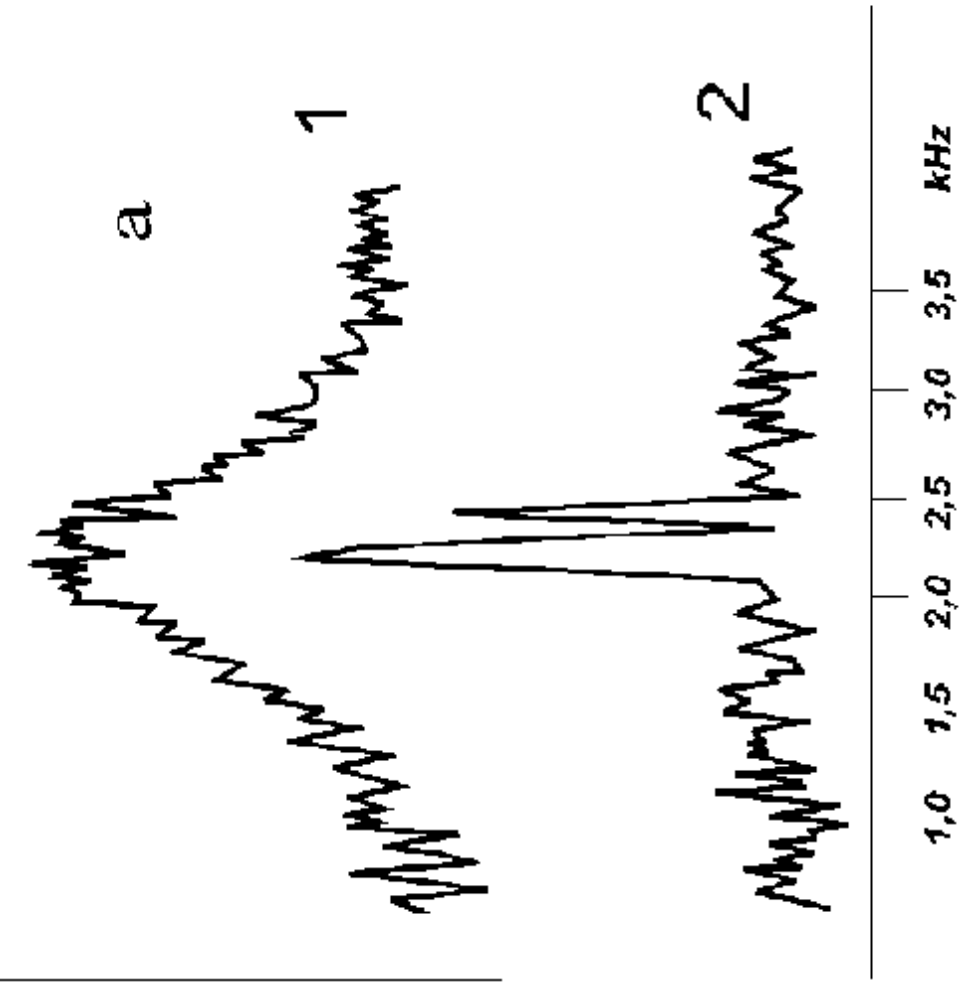
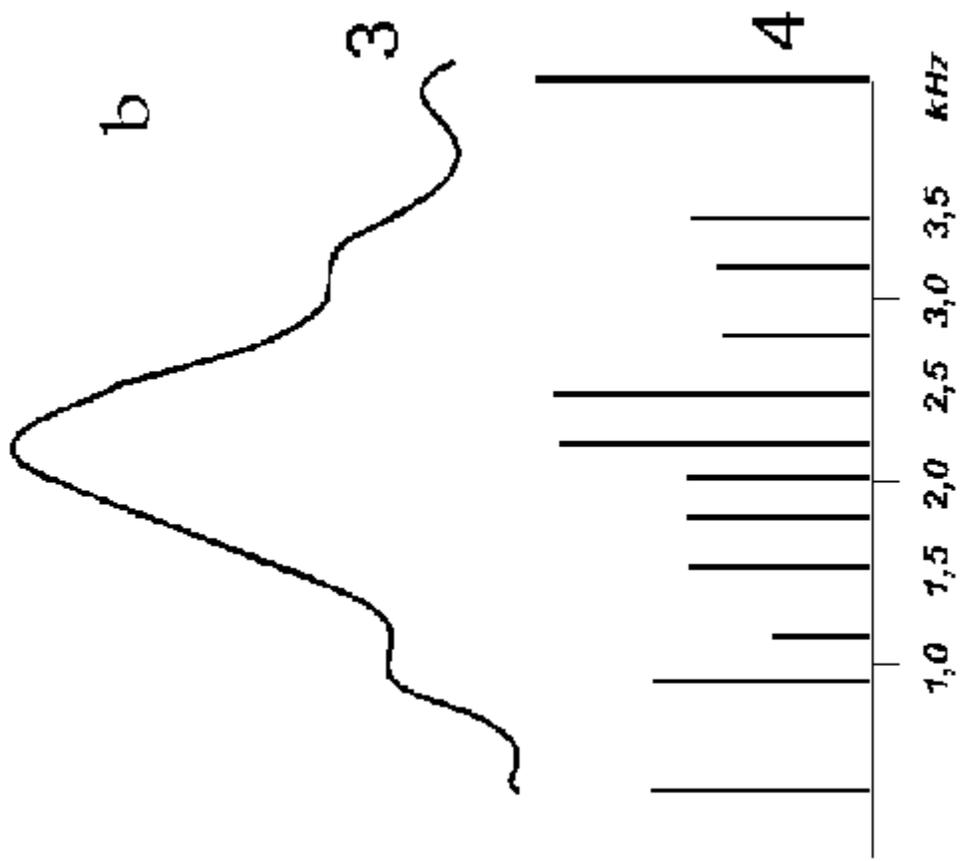
\end{document}